\documentstyle[12pt]{article}
\begin{document}
\begin{center}
\Large
{\bf
SUPERFIELD QUANTIZATION OF GENERAL GAUGE THEORIES}

\vspace{0.5cm}

P.M.LAVROV\footnote{E-mail address: lavrov@tspi.tomsk.su}\\
Tomsk State Pedagogical University, Tomsk, 634041, Russia
\end{center}
\vspace{.5cm}
\large
\begin{quotation}
\normalsize
\noindent{\bf Abstract}

A superfield version on superspace $(x^\mu,\theta^a)$ is proposed
for the $Sp(2)$-- covariant Lagrangian quantization of general
gauge theories. The BRST- and antiBRST- transformations are realized
on superfields as supertranslations in the $\theta^a$-- directions.
A new (geometric) interpretation of the Ward identities in the
quantum gauge theory is given.
\end{quotation}
\section{INTRODUCTION}
\hspace*{\parindent}
Recently, it has been realized [1-8] that BRST-antiBRST
invariance can be used as fundamental principle in the construction of quantum
gauge theories. This principle has been applied in [1-3] to
quantization of dynamical systems with constraints in the Hamiltonian
formalism. The $Sp(2)$-- covariant Lagrangian quantization of general
gauge theories has been developed in [4-6] on the basis of the BRST-
antiBRST symmetry principle. Note that in [7] a slightly different
(but completely equivalent to [4,5]) procedure of the $Sp(2)$--
covariant Lagrangian quantization was suggested. A completely anticanonical
form of the $Sp(2)$-- covariant Lagrangian quantization scheme (triplectic
quantization in terms of [9]) has been given in [8].

Geometric contents of the BRST- and antiBRST- transformations
for the Yang-Mills theories as supertranslations in superspace
along additional (Grassmann) coordinates has been realized quite
a long time ago [10-15], but there has not been found any satisfactory
superfield description of the quantization procedure. Moreover, the
crucial point of the superfield descriptions [10-15] was the manifest
structure of the Yang-Mills  theory action, the problem of treatment
of arbitrary gauge theories has remained unsolved.

The main purpose of this paper is to give the superfield quantization
rules for general gauge theories, based on the principle of invariance
under the BRST- and antiBRST- transformations acting in superspace
as supertranslations along the additional (with respect to the
space-time coordinates) variables.

The paper is organized as follows. In section 2 the basic objects
of the superfield quantization are introduced: superfields $\Phi^A(
\theta)$, supersources $\bar{\Phi}_A(\theta)$, super-antibrackets
$(\;,\;)^a$, operators $\Delta^a$, $V^a$; besides some properties
of the objects introduced are studied. In section 3 the superfield
quantization rules for general gauge theories are formulated. The
generating functional of Green functions is constructed as a functional
depending on supersources $\bar{\Phi}_A$. The superfield form of the
BRST- and antiBRST- transformations is found. Independence of the
$S$-- matrix on a choice of the gauge is proved. The Ward identities
in terms of superfields are derived. In section 4 a connection
between the scheme in question and the triplectic [8,9] as well as
the $Sp(2)$-- covariant [4-6] quantization methods is established,
the latter being some particular cases of the superfield quantization,
corresponding to special choices of both the solutions of superfield
generating equations and the gauge. In section 5 the concluding
remarks are made.

In the paper the condensed notations by DeWitt [16] are used. The
Grassmann parity of a quantity $A$ is denoted $\varepsilon(A)$.
Derivatives with respect to (super)fields are understood as the
right-hand and those with respect to antifields and (super)sources
as the left-hand ones. The left derivatives with respect to (super)fields
are labelled $"{\it l}"$: $\delta_{\it l}/ \delta\phi^A$. Scalar
anticommuting coordinates $\theta^a$ form an $Sp(2)$-- doublet.
Lowering the $Sp(2)$-- indices is given by the rule $\theta_a=
\varepsilon_{ab}\theta^b$. Derivatives with respect to $\theta^a$
are understood as the left-hand ones. Integration over $\theta^a$
is given by
$$
\int d^2\theta=0,\;\;\int d^2\theta\;\theta^a=0,\;\;\int d^2\theta\;
\theta^a\theta^b=\varepsilon^{ab}.
$$
For any function $f(\theta)$ the equalities hold
$$
\int d^2\theta\;\frac{\partial f(\theta)}{\partial \theta^a}=0,
$$
Any function of $\theta$ can be represented in the form
$$
f(\theta)=f_o+f_a\theta^a+\frac{1}{2}f_3\theta_a\theta^a.
$$

\section{MAIN DEFINITIONS}
\hspace*{\parindent}
Let us introduce a superspace with coordinates $(x^\mu,\theta^a)$,
where $x^\mu$ $(\mu=0,1,...,d-1)$ are the space-time coordinates
and $\theta^a$ $(a=1,2)$ are anticommuting scalar coordinates. Let
$\Phi^A(\theta)$, $\varepsilon(\Phi^A(\theta))\equiv \varepsilon_A$
be a set of superfields with the following restriction
$$
\left.\Phi^A(\theta)\right|_{\theta=0}=\phi^A, \eqno(1)
$$
where $\phi^A$ are the fields of configuration space in the $Sp(2)$--
covariant Lagrangian quantization [4-6]. We associate with each
superfield $\Phi^A(\theta)$ one supersource $\bar{\Phi}_A(\theta)$
of {\it the same} Grassmann parity,
$$
\varepsilon(\bar{\Phi}_A(\theta))=\varepsilon_A. \eqno(2)
$$
In terms of supervariables $\Phi^A(\theta),\bar{\Phi}_A(\theta)$
we define for any functionals $F=F(\Phi,\bar{\Phi})$,
$G=G(\Phi,\bar{\Phi})$ the super-antibrackets
$$
(F,G)^a=\int d^2\theta\left\{\frac{\delta F}{\delta\Phi^A(\theta)}
(-1)^{\varepsilon_A+1}\frac{\partial}{\partial\theta_a}\frac{\delta G}
{\delta\bar{\Phi}_A(\theta)}-\right.
$$
$$
 (F\leftrightarrow G)(-1)^{(\varepsilon (F)+1)
(\varepsilon (G)+1)}\Bigg\} \eqno(3)
$$
with the evident property of generalized antisymmetry
$$
(F,G)^a=-(-1)^{(\varepsilon (F)+1)(\varepsilon (G)+1)}(G,F)^a. \eqno(4)
$$
The super-antibrackets (3) satisfy the generalized Jacobi identities
$$
((F,G)^{\{a},H)^{b\}}(-1)^{(\varepsilon (F)+1)(\varepsilon (H)+1)}+cycle
(F,G,H)\equiv 0, \eqno(5)
$$
where the curly bracket denotes symmetrization of $a$ and $b$.

Let us introduce the operators $\Delta^a$ and $V^a$ by the rule
$$
\Delta^a=-\int d^2\theta \frac {\delta_{\it l}}{\delta\Phi^A(\theta)}
\frac {\partial}{\partial \theta_a}\frac
{\delta}{\delta\bar{\Phi}_A(\theta)}, \eqno(6)
$$
$$
 V^a=\int d^2\theta\left(\frac {\partial\bar{\Phi}_A(\theta)}{\partial
\theta_a}\frac {\delta}{\delta\bar{\Phi}_A(\theta)}+\frac {\partial
\Phi^A(\theta)}{\partial\theta_a}\frac {\delta_l}{\delta\Phi^A(\theta)}
\right). \eqno(7)
$$
One may easily check that the following properties hold
$$
\Delta^{\{a}\Delta^{b\}}=0, \eqno(8)
$$
$$
V^{\{a}V^{b\}}=0, \eqno(9)
$$
$$
\Delta^aV^b+V^b\Delta^a=0. \eqno(10)
$$
The action of the operators $\Delta^a$ and $V^a$ upon the super-
antibrackets is given by the relations
$$
\Delta^{\{a}(F,G)^{b\}}=(\Delta^{\{a}F,G)^{b\}}-(F,\Delta^{\{a}G)^{b\}}
(-1)^{\varepsilon(F)}, \eqno(11)
$$
$$
V^a(F,G)^b=(V^aF,G)^b-(F,V^aG)^b(-1)^{\varepsilon(F)}. \eqno(12 )
$$
It is also convinient to introduce the extended operators $\bar{\Delta}^a$
$$
\bar{\Delta}^a=\Delta^a+\frac{i}{\hbar}V^a. \eqno(13)
$$
These operators satisfy the relations
$$
\bar{\Delta}^{\{a}\bar{\Delta}^{b\}}=0, \eqno(14)
$$
which follow from (8)-(10).

\section{SUPERFIELD QUANTIZATION}
\hspace*{\parindent}
The basic object of superfield quantization is the quantum action
$W=W(\Phi,\bar{\Phi})$. We require that $W$ be a solution to the
following generating equations (quantum master equations)
$$
\bar{\Delta}^a\exp\left\{\frac{i}{\hbar}W\right\}=0 \eqno(15)
$$
or equivalently
$$
\frac{1}{2}(W,W)^a+V^aW=i\hbar\Delta^aW. \eqno(16)
$$

The generating functional of the Green functions $Z=Z(\bar{\Phi})$
for superfields $\Phi^A(\theta)$ we define as
$$
Z(\bar{\Phi})=\int[d\Phi'][d\bar{\Phi'}]\rho (\bar{\Phi'})\exp \left\{
\frac{i}{\hbar}\bigg[W(\Phi',\bar{\Phi'})-\right.
$$
$$
-\left.\frac{1}{2}\varepsilon_{ab}
V'^aV'^bF(\Phi')-\bar{\Phi}\Phi'\bigg]\right\},\eqno(17)
$$
where $F(\Phi)$ is the boson gauge functional, $\rho(\bar{\Phi})$
is the weight functional having the form of functional $\delta$--
function
$$
\rho(\bar{\Phi})=\delta\left(\int d^2\theta\;\bar{\Phi}(\theta)\right),
\eqno(18)
$$
and the notation
$$
\bar{\Phi}\Phi\equiv\int d^2\theta\;\bar{\Phi}_A(\theta)\Phi^A(\theta)
\eqno(19)
$$
is used.

The introduced above generating functional (17) possesses two important
properties. Firstly, the integrand in (17) for $\bar{\Phi}=0$
is invariant under the transformations of global supersymmetry
$$
\delta\Phi^A(\theta)=\mu_a\frac{\partial\Phi^A(\theta)}{\partial\theta_a},
\quad\varepsilon(\mu_a)=1, \eqno(20)
$$
$$
\delta\bar{\Phi}_A(\theta)=\mu_a\frac{\partial\bar{\Phi}_A(\theta)}
{\partial\theta_a}+\mu_a\frac{\partial}{\partial\theta_a}
\frac{\delta W}{\delta\Phi^A(\theta)} \eqno(21)
$$
on account of the generating equations (16) and the invariance of weight
functional (18) under transformations (21)
$$
\delta\rho(\bar{\Phi})=0. \eqno(22)
$$
In (20),(21) $\mu_a$ is a $Sp(2)$-- doublet of the constant anticommuting
Grassmann parameters. Secondly, the vacuum functional $Z(0)$ does not
depend on choice of the gauge boson $F$ within the proposed superfield
scheme (16)-(18). Indeed, suppose $Z_F\equiv Z(0)$. We shall change
the gauge $F(\Phi)\rightarrow F(\Phi)+\delta F(\Phi)$. In the functional
integral for $Z_{F+\delta F}$ we make the change of variables (20),(21)
choosing for the parameters $\mu_a$
$$
\mu_a=-\frac{i}{2\hbar}\varepsilon_{ab}V^b\delta F(\Phi). \eqno(23)
$$
Taking into account (9),(16) and (22) we find that
$$
Z_{F+\delta F}=Z_F \eqno (24)
$$
and, hence, the $S$-- matrix is gauge-invariant.

The transformations (20), (21) realize the BRST- and antiBRST- symmetry
in the superfield approach to quantum gauge theory. Allowing for (3),(7)
one can rewrite these transformations in the form
$$
\delta\Phi^A(\theta)=\mu_aV^a\Phi^A(\theta), \eqno(25)
$$
$$
\delta\bar{\Phi}_A(\theta)=\mu_aV^a\bar{\Phi}_A(\theta)+
\mu_a(W,\bar{\Phi}_A(\theta))^a. \eqno(26)
$$
{}From (25), (26) we conclude that the operators $V^a$ (7) can be
considered as generators of supertranslations in the $\theta^a$--
directions on supervariables $\Phi^A(\theta),\bar{\Phi}_A(\theta)$.
This gives a geometric interpretation of the BRST- and antiBRST-
symmetry for arbitrary gauge theory.

Invariance of vacuum functional $Z(0)$ under the BRST- and antiBRST-
transfomations leads to the presence of gauge Ward identities. Let us
consider the derivation of these identities. To do this we shall use
the standard assumptions on functional integral properties, in particular,
$$
\int [d\Phi][d\bar{\Phi}]\rho(\bar{\Phi})\;\frac{\delta F(\Phi,\bar{\Phi})}
{\delta\Phi}=0,\;\int d[\Phi][d\bar{\Phi}]\rho(\bar{\Phi})\;
\frac{\delta F(\Phi,\bar{\Phi})}{\delta \bar{\Phi}}=0. \eqno(27)
$$
Taking into account the explicit form of the operators $\bar{\Delta}^a$
and (27) we have the following equalities
$$
\phantom{I}
\begin{array}{c}
\displaystyle\int [d\Phi'][d\bar{\Phi}']\rho(\bar{\Phi}')\bar\Delta'^a
\exp\left\{\frac{\displaystyle i}{\displaystyle\hbar}
[ W(\Phi',\bar{\Phi}')- \right.\\
\phantom{o} \\
-\left.\frac{\displaystyle1}{\displaystyle2}
\varepsilon_{ab}V'^aV'^bF(\Phi')-\bar{\Phi}\Phi']\right\}=0
\end{array}
\eqno(28)
$$
Let us act on exponential by the operators $\bar{\Delta}^a$ and take
into account (9), (10), (16). We obtain
$$
V^aZ(\bar{\Phi})=0.\eqno(29)
$$
Equations (29) represent the superfield form of Ward identities for
generating functional of Green functions.

Let us now consider the Legendre transformation of $\ln Z(\bar{\Phi})$
with respect to the supersources $\bar{\Phi}_A(\theta)$. For this
purpose we define superfields $\Phi^A(\theta)$ by the relations
$$
\Phi^A(\theta)=-\frac{\hbar}{i} \frac{\delta\ln Z(\bar\Phi)}
{\delta\bar{\Phi}_A(\theta)} \eqno(30)
$$
and introduce the functional $\Gamma(\Phi)$ by the rule
$$
\Gamma(\Phi)=\frac{\hbar}{i}\;\ln Z(\bar{\Phi})+\bar{\Phi}\Phi, \eqno(31)
$$
so that
$$
\frac{\delta\Gamma}{\delta\Phi^A(\theta)}=\bar{\Phi}_A(\theta). \eqno(32)
$$
$\Gamma(\Phi)$ is the vertex functions generating functional. Rewriting
the Ward identities (29) for the vertex functios generating functional,
we have the relations
$$
V^a\Gamma(\Phi)=0 \eqno(33)
$$
as the superfield form of the Ward identities for $\Gamma(\Phi)$.

{}From (29), (33) one can establish a new (geometric) interpretation
of the Ward identities in quantum gauge theory. Indeed, the Ward
identities express the invariance of the generating functionals
$Z$ and $\Gamma$ under supertranslations in the $\theta^a$--
directions.

\section{CONNECTION WITH THE TRIPLECTIC AND
         {\sl Sp(2)\/}--COVARIANT QUANTIZATIONS}
\hspace*{\parindent}
It is useful to compare the suggested superfield quantization for
general gauge theories with the $Sp(2)$-- covariant Lagrangian
quantization [4-6] and the triplectic one [8]. To this end we
present, first of all, the matter of developed superfield quantization
scheme in the component form.

For supervariables $\Phi^A(\theta)$ and $\bar{\Phi}_A(\theta)$
we shall use the following notations in the $\theta^a$-- expansions
$$
\Phi^A(\theta)=\phi^A+\pi^{Aa}\theta_a+\frac{1}{2}\lambda^A\theta_a
\theta^a,\eqno(34)
$$
$$
\bar{\Phi}_A(\theta)=\bar{\phi}_A-\theta^a\phi^*_{Aa}-\frac{1}{2}\theta_a
\theta^aJ_A.\eqno(35)
$$
Then the operators $\Delta^a$ (6), $V^a$ (7) and super-antibrackets (3)
have the form
$$
(F,G)^a=\frac{\delta F}{\delta\phi^A}\;\frac{\delta G}{\delta\phi^*_{Aa}}
+\varepsilon^{ab}\frac{\delta F}{\delta\pi^{Ab}}\frac{\delta
G}{\delta\bar{\Phi}_A} -
$$
$$
(F\leftrightarrow G)\;(-1)^{(\varepsilon(F)+1)(\varepsilon(G)+1)},
\eqno(36)
$$
$$
\Delta^a=(-1)^{\varepsilon_A}\frac{\delta_{\it l}}{\delta\phi^A}\;
\frac{\delta}{\delta\phi^*_{Aa}}+(-1)^{\varepsilon_A+1}\varepsilon^{ab}
\frac{\delta_{\it l}}{\delta\pi^{Ab}}\;\frac{\delta}{\delta\bar{\phi}_A},
\eqno(37)
$$
$$
V^a=\varepsilon^{ab}\phi^*_{Ab}\frac{\delta}{\delta\bar{\phi}_A}-
(-1)^{\varepsilon_A}\pi^{Aa}\frac{\delta_{\it l}}{\delta\phi^A}+
$$
$$
+(-1)^{\varepsilon_A}\varepsilon^{ab}\lambda^A\frac{\delta_{\it l}}{\delta
\pi^{Ab}}-J_A\frac{\delta}{\delta\phi^*_{Aa}}. \eqno(38)
$$
In the component form the gauge fixing action reads
$$
\phantom{o}
\begin{array}{c}
\frac{\displaystyle1}{\displaystyle2}\varepsilon_{ab}V^aV^bF(\Phi)=
\frac{\displaystyle1}{\displaystyle2}\varepsilon_{ab}
\pi^{Aa}\frac{\displaystyle\delta^2F}{\displaystyle\delta\phi^A\delta\phi^B}
\pi^{Bb}+
\frac{\displaystyle\delta F}{\displaystyle\delta\phi^A}\lambda^A- \\
\phantom{o} \\
-\frac{\displaystyle1}{\displaystyle2}\varepsilon^{ab}
\lambda^A\frac{\displaystyle\delta^2F}
{\displaystyle\delta\pi^{Aa}\delta\pi^{Bb}}\lambda^B+\pi^{Aa}
\frac{\displaystyle\delta^2F}{\displaystyle\delta\phi^A\delta\pi^{Ba}}
\lambda^B,
\end{array}
\leqno(39)
$$
For the functional $\bar{\Phi}\Phi$ in (17) we have
$$
\bar{\Phi}\Phi=\bar{\phi}_A\lambda^A+\phi^*_{Aa}\pi^{Aa}-J_A\phi^A.
\eqno(40)
$$
Integration measure
$$
[d\Phi][d\bar{\Phi}]\rho(\bar{\Phi})=[d\phi][d\phi^*][d\pi][d\bar{\phi}]
[d\lambda][dJ]\delta(J) \eqno(41)
$$
coincides, in fact, with measures in functional integrals of papers
[4-6] and [8].

Let us now turn ourselves to the generating equations (16) of the
superfield formalism. By virtue of (36)-(38) we can see that the
variables $\lambda^A$ and $J_A$ enter the equations (16) in a very
special manner, being only contained in the operators $V^a$ as linear
multipliers at the differential operators. We shall make use of this
special point. To this end, we note that the functional $\bar{\Phi}\Phi$
is linear with respect to $\lambda^A$, $J_A$ and satisfies both the
equations
$$
V^a\bar{\Phi}\Phi=0 \eqno(41)
$$
and (16) as a whole. This enables one to seek solutions of the equations
(16) in the form
$$
W(\Phi,\bar{\Phi})=S(\Phi,\bar{\Phi})+\bar{\Phi}\Phi, \eqno(42)
$$
with the functional $S(\Phi,\bar{\Phi})$ subjected to the conditions
$$
\int d^2\theta\frac{1}{2}\theta_a\theta^a\frac{\delta S}{\delta\Phi^A
(\theta)}=\frac{\delta S}{\delta\lambda^A}=0,
$$
$$
\int d^2\theta\frac{1}{2}\theta_a\theta^a\frac{\delta S}{\delta
\bar{\Phi}_A(\theta)}=-\frac{\delta S}{\delta J_A}=0.\eqno(43)
$$
Taking (43) and (36)-(38) into account, we find that the functional
$S$ is exactly a solution to the generating equations of paper [8].
Consequently, the special choice of solutions to the equations (16)
in the form (42),(43) with $\bar{\Phi}=0$ leads to the vacuum
functional of paper [14], corresponding to the special choice of the
gauge (note that in [14], apart from a particular choice of gauge
fixing, the general approach to solution of this problem is studied).

One can also readily establish a connection with the $Sp(2)$-- covariant
quantization of papers [4-6]. To do this, suffice it to require, in
addition, that the functional $S$ be independent on the variables
$\pi^{Aa}$, i.e.
$$
\frac{\delta S}{\delta \pi^{Aa}}=0,
$$
choosing the gauge fixing functional $F$ as to depend on the variables
$\phi^A$ only: $F=F(\phi)$. Then we find in (17) the exact form of the
generating functional of Green functions in the $Sp(2)$-- covariant
quantization of papers [4-6].

The form (29) of the Ward identities for $Z(\bar{\Phi})$, rewritten in
terms of the components
$$
J_A\frac{\delta Z}{\delta \phi^{*}_{Aa}}-\varepsilon^{ab}\phi^{*}_{Ab}
\frac{\delta Z}{\delta\bar\phi_A}=0
\eqno(44)
$$
coincides with the one, derived in [4].

The component form of the definition (30) for $\Phi^A (\theta)$ reads
$$
\phi^A=\frac{\hbar}{i}\frac{\delta \ln Z}{\delta J_A}, \quad
\pi^{Aa}=\frac{\hbar}{i}\frac{\delta \ln Z}{\delta \phi^{*}_{Aa}}, \quad
\lambda^A=\frac{\hbar}{i}\frac{\delta \ln Z}{\delta\bar\phi_A},
$$

Given this, the equations (32) can be written as
$$
\frac{\delta\Gamma}{\delta\phi^A}=-J_A, \quad
\frac{\delta\Gamma}{\delta\pi^{Aa}}=\phi^{*}_{Aa}, \quad
\frac{\delta\Gamma}{\delta\lambda^A}=\bar\phi_A.
$$
For the Ward identities (34) we have the following equations
$$
\pi^{Aa}\frac{\delta\Gamma}{\delta\phi^A}+\varepsilon^{ab}\lambda^A
\frac{\delta\Gamma}{\delta\pi^{Ab}}=0
\eqno(45)
$$
The form (45) of the Ward identities for the generating functional of
vertex functions $\Gamma$ differs from the one, obtained in [4]. The
difference is due to the fact that in [4] the transition from the Ward
identities for $Z$ (coinciding, formally, with (44)) to the ones for
$\Gamma$ was made by the Legendre transformation with respect to the
variables $J_A$ only, the variables $\phi^*_{Aa}$, $\bar{\phi}_A$ being
inert.

\section{DISCUSSION}
\hspace*{\parindent}
In this paper a closed superfield representation of the $Sp(2)$--
covariant quantization [4-6] for gauge theories is found. One of the
basic objects of the method in question is a superfield $\Phi^A (\theta)$
subject to the boundary condition $\Phi^A(\theta)|{}_{\theta=0}=\phi^A$,
where $\phi^A$ is the set of fields of configuration space in the
$Sp(2)$--covariant quantization formalism [4-6]. To each superfield $\Phi^A
(\theta)$ there is assigned the corresponding supersource $\bar{\Phi}_A
(\theta)$ with the same Grassmann parity. The variables belonging to the
complete set of the $Sp(2)$-- covariant quantization method arise in the
superfield formalism quite naturally as the corresponding components of
all the supervariables $\Phi^A (\theta)$, $\bar{\Phi_A}(\theta)$ (see (34),
(35)).

The BRST-antiBRST symmetry transformations (20),(21) (or, equivalently
(25),(26)) for arbitrary gauge theories have within the superfield
formalism a clear geometric contents as they are realized as
supertranslations in superspace $(x^\mu, \theta^a)$ along the Grassmann
coordinates $\theta^a$.

The superfield description permits one to have a new look at the Ward
identities in the quantum theory of gauge fields, thus revealing their
geometric contents. Indeed, the identities (29) and (33) for both the
generating functional of Green functions $Z=Z(\bar{\Phi})$ and the
generating functional of vertex functions $\Gamma =\Gamma (\Phi)$ are
nothing but the statements that $Z$ and $\Gamma$ are invariant under
supertranslations in superspace. Simultaneously, the role and geometric
origin of the operators $V^a$, realizing the supertranslations in terms
of the variables $\Phi^A(\theta)$, $\bar{\Phi}_A(\theta)$ are revealed.

The superfield description permits a universal outlook on the quantization
scheme for general gauge theories, developed in papers [4-6] and [8,9];
the latter are the particular cases of the superfield description as they
correspond to certain special choices of both the solutions to the
generating equations (16) and the gauge.\\

\vspace{2mm}
\noindent{\bf Acknowledgement}
\vspace{.5mm}

The author is grateful to I.V.Tyutin and V.I.Mudruk for helpful
disscussions. This work has been supported in part by grant RI~1300 from
the International Science Foundation and by the Russian Fund for
Fundamental studies under grant number 94-02-03234.
\newpage
\begin{center}
{\bf References}
\end{center}
\medskip
\begin{itemize}

\item[{\hfill [1]}]
Batalin I.A., Lavrov P.M., Tyutin I.V.,
J.~Math. Phys.
31 (1990) 6.

\item[{\hfill [2]}]
Batalin I.A., Lavrov P.M., Tyutin I.V.,
J.~Math. Phys.
31 (1990) 2708.

\item[{\hfill [3]}]
Batalin I.A., Lavrov P.M., Tyutin I.V.,
Int. J.~Mod. Phys.
6 (1991) 3599.

\item[{\hfill [4]}]
Batalin I.A., Lavrov P.M., Tyutin I.V.,
J.~Math. Phys.
31 (1990) 1487.

\item[{\hfill [5]}]
Batalin I.A., Lavrov P.M., Tyutin I.V.,
J.~Math. Phys.
32 (1991) 532.

\item[{\hfill [6]}]
Batalin I.A., Lavrov P.M., Tyutin I.V.,
J.~Math.~Phys.
32 (1991) 2513.

\item[{\hfill [7]}]
Hull C.M.,
Mod.~Phys. Lett.
A5 (1991) 1871.

\item[{\hfill [8]}]
Batalin I., Marnelius R.,
G\"{o}teborg preprint ITP 94--30 (1995).

\item[{\hfill [9]}]
Batalin I.A., Marnelius R., Semikhatov A.M.,
G\"{o}teborg preprint ITP 94--31 (1995).

\item[{\hfill [10]}]
Bonora L., Tonin M.,
Phys. Lett.
B98 (1981) 48.

\item[{\hfill [11]}]
Bonora L., Pasti P., Tonin M.,
J. Math. Phys.
23 (1982) 839.

\item[{\hfill [12]}]
Hoyos J. Quiros M., Mittelbrunn J.R., De Urries F.,
Nucl. Phys.
B218 (1983) 159.

\item[{\hfill [13]}]
Baulieu L.,
Nucl. Phys.
B218 (1983) 157.

\item[{\hfill [14]}]
Baulieu L.,
Phys. Rep.
129 (1985) 1.

\item[{\hfill [15]}]
Hull C.M., Spence B., Vazquez-Bello J.L.,
Nucl. Phys.
B348 (1991) 108.

\item[{\hfill [16]}]
De Witt B.S.,
Dynamical Theory of Groups and Fields (Gordon and Breach, 1965).
\end{itemize}

\end{document}